\documentclass[aps,onecolumn,showpacs,preprintnumbers]{revtex4-1}

\usepackage{graphicx}  
\usepackage{subfigure}
\usepackage{multirow}

\usepackage{fancyhdr}
\usepackage{longtable}
\usepackage{parskip}
\usepackage[T1]{fontenc}
\usepackage{dcolumn}   
\usepackage{color}
\usepackage{bm}        
\usepackage{amsfonts}  
\usepackage{amsmath}   
\usepackage{amssymb}   
\usepackage[utf8]{inputenc}
\usepackage{physics}

\begin{document}

\title{Incompatibility of optimized protection of entanglement and teleportation fidelity in the presence of decoherence}
\author{Priyanka Chowdhury}
\thanks{priyanka.chowdhury@basirhatcollege.org}
\affiliation{\it Department of Physics, Basirhat College, Basirhat, 743412, India}


\begin{abstract}
\noindent Entanglement is the key success of teleporting an unknown quantum state with fidelity higher than  classical limit. In the presence of decoherence, entanglement decreases with the strength of interaction between quantum systems and the environment. As a result, teleportation fidelity (TF) decreases. The technique of weak measurement and its reversal help to protect entanglement in the presence of amplitude-damping decoherence. In this work, we have shown that the optimal protection of entanglement does not optimize TF. More specifically, when one of the systems interacts with the environment, optimized TF requires higher strength of reverse weak measurement than optimized entanglement protection. The success probability of optimal protection indicates that higher form of nonlocal correlation plays the key role in optimizing TF. Interestingly, when both systems interact with the environment, optimization of entanglement implies optimization of TF. Therefore, the resources of quantum teleportation along with entanglement need to be explored.
\\\\
\textbf{Keywords :} Teleportation; Entanglement; Bell nonlocality; Weak measurement; Decoherence.\\

\end{abstract}



\maketitle

\section{Introduction}

Quantum teleportation (QT)~\cite{BBCJP_93,BPMEWZ_97, DBMHP_98, MRTZG_03, Ren_al_17} is an information processing task where the sender, say, Alice would like to communicate an unknown quantum state to a receiver, Bob located at a distant position without physically sending the quantum system. It has been shown that quantum correlation plays a significant role in different information processing tasks than classical correlation~\cite{Book_NC}. In the case of QT, entanglement~\cite{Ent_rev} provides the efficiency of QT (measured by the fidelity of the teleported state with the given unknown state) higher than $2/3$, which is the upper bound of the teleportation fidelity (TF) obtained using classical correlation\cite{PW_91, MP_95}. In other words, entanglement becomes necessary to obtain TF in the non-classical region, i.e., higher than $2/3$. By sharing a maximally entangled state, Alice can perform QT with fidelity of unity by communicating $2$-cbits (classical bits) to Bob.

Recent developments include the increment of distance of QT~\cite{MRTZG_03, JY_12, XSM_12, Ren_al_17}. Earth-to-satellite QT~\cite{Ren_al_17}  opens the feasibility of global quantum communication. The efficiency of long-distance QT is lower than unity; e.g., the average TF achieved in earth-to-satellite QT is $0.80 \pm 0.01$ ~\cite{Ren_al_17}. The decrement of fidelity occurs due to the effect of decoherence. In the presence of decoherence, the quantum correlation, e.g., entanglement, decreases gradually with the strength of interaction between quantum systems and the environment~\cite{YE_09, SMAHWRD_08, CCAA_12, AMHSWRD_07, PCHLMK_19}. As a result, TF also decreases when the strength of decoherence increases~\cite{OLL_02, PCPS_07, PM_13}. Note that entanglement can be generated under certain circumstances when systems collectively interact with the common environment~\cite{PH_02, DB_02, KLAK_02}. It has been shown that decoherence can activate TF, i.e., enhancement of TF from classical region to non-classical region without increment of entanglement~\cite{BHHH_00, Bandyopadhyay_02,NDDA_18, QTYW_14}.

Several techniques have been proposed to minimize the effect of decoherence on quantum correlation~\cite{PM_13, PH_02, DB_02, KLAK_02, KU_99,KCRK_09,  LJKK_11, KLKK_12, SP_95, PHBK_99, NV_11, PCHLKM_19, LHK_15}. For example, the effect of environmental interaction modelled by the amplitude damping channel (ADC) can be reduced using the technique of weak measurement and reverse weak measurement(WMRWM)~\cite{PM_13, KU_99, KCRK_09, LJKK_11, KLKK_12, QTYW_14, HZC_23, SKB_23}. Different forms of quantum correlation, e.g, coherence\cite{KCRK_09, LJKK_11}, entanglement~\cite{KLKK_12}, QT~\cite{PM_13} can be protected in the presence of decoherence modelled by ADC. More specifically, using the technique of WMRWM quantum correlation can be useful at higher strength of decoherence than without using it. But, the technique fails to protect QT when entanglement sudden death (ESD)~\cite{YE_09,SMAHWRD_08, AMHSWRD_07, PCHLMK_19} occurs.

The technique of WMRWM increases the amount of entanglement\cite{KLKK_12} with respect to (w.r.t.) the strength of weak measurement, and as a result, TF is also enhanced~\cite{PM_13}. Here, optimal protection of entanglement has been obtained from its maximization w.r.t. the strength of reverse weak measurement. As entanglement is the resource of QT, it seems that maximization of entanglement also maximizes the TF. But, in this work, we have shown that the above is not always true, i.e., optimization of entanglement does not optimize the teleportation fidelity. For this purpose, two qubits have been prepared in the maximally entangled state, and they are allowed to interact with the environment via ADC. We have studied the optimal protection of both entanglement and TF w.r.t. the strength of reverse weak measurement. We have found that when one of the qubits has been affected by ADC, optimized TF requires higher strength of reverse weak measurement than optimized entanglement. In this case, the success probability associated with the technique of WMRWM reveals that optimized TF requires stronger form of nonlocal correlation than entanglement. Interestingly, when both qubits interact with the environment, optimization of entanglement implies optimization of TF.


\section{Environmental interaction modelled by ADC and characterization of quantum correlation}

There are different theoretical models describing the interaction between quantum systems and the environment~\cite{Book_NC}. In the present work, we have considered the decoherence modelled by ADC. According to this model, when the system, say, $i$th $(i\in\{1,\,2\})$ qubit, is in the ground energy state ($|0\rangle_i$), it is unaffected by the environment. But, when the system is in the exited energy state ($|1\rangle_i$), the system jumps to the state $|0\rangle_i$ with a probability~$D_i$ by spontaneously emitting a photon, and it remains in the state $|1\rangle_i$ with probability $\overline{D}_i=1-D_i$. The interaction can be written as
\begin{eqnarray}
|0\rangle_i |0\rangle_E &&\longrightarrow |0\rangle_i |0\rangle_E, \nonumber \\
|1\rangle_i |0\rangle_E && \longrightarrow \sqrt{\overline{D}_i} |1\rangle_i|0\rangle_E + \sqrt{D_i} |0\rangle_i|1\rangle_E,
\label{ADC_Map}
\end{eqnarray}
where $|0\rangle_E$ is the initial state of the environment. 
The above interaction~(\ref{ADC_Map}) can be expressed as a map $\Lambda$,
\begin{eqnarray}
\Lambda(\rho_i) = \mathcal{D}_{i,0} \, \rho_i \,  \mathcal{D}_{i,0}^\dagger +  \mathcal{D}_{i,1} \, \rho_i \,  \mathcal{D}_{i,1}^\dagger,
\label{ADC}
\end{eqnarray}
where the Kraus operators $ \mathcal{D}_{i,\,j}$ are given by
\begin{eqnarray}
 \mathcal{D}_{i,0} &= &
\begin{pmatrix}
1 & 0\\
0 & \sqrt{\overline{D}_i}
\end{pmatrix}, 
\nonumber \\
 \mathcal{D}_{i,1} &=& 
\begin{pmatrix}
0 & \sqrt{D_i}\\
0 & 0
\end{pmatrix},
\end{eqnarray}
and $\sum_{j=0}^1  \mathcal{D}_{i,j}  \mathcal{D}_{i,j}^\dagger = \rm{I}$. 

To study the effect of decoherence, concurrence and TF have been considered to measure entanglement and efficiency of QT, respectively. The entanglement of a given two-qubit state $\rho_{AB}$ can be quantified by the concurrence~\cite{Wootters_98},
\begin{eqnarray}
\mathcal{C}(\rho_{AB}) = \max\left[0, \sqrt{\lambda_1}-\sqrt{\lambda_2}-\sqrt{\lambda_3}-\sqrt{\lambda_4} \right],
\label{Concurrence}
\end{eqnarray}
where $\lambda_i$'s are eigenvalues of the matrix $\rho_{AB} \tilde{\rho}_{AB}$ in descending order. Here, $\tilde{\rho}_{AB} = (\sigma_y\otimes \sigma_y)\rho_{AB}^*(\sigma_y\otimes\sigma_y)$, where the superscript $``*"$ represents the complex conjugate. The value of $\mathcal{C}$ is bounded by $[0,\,1]$, where the lower bound $0$ and upper bound $1$ correspond to separable and maximally entangled states, respectively.

The maximum attainable TF of the shared entangled state $\rho_{AB}$ is given by~\cite{HHH_99}
\begin{eqnarray}
F(\rho_{AB}) = \frac{2 f(\rho_{AB}) + 1}{3},
\label{TF}
\end{eqnarray}
where the fully entanglement fraction (FEF) $f(\rho_{AB})$ is calculated as~\cite{BVSW_96}
\begin{eqnarray}
f(\rho_{AB}) = \max_{|\phi\rangle \in \text{MES}} \langle \phi|\rho_{AB}|\phi \rangle,
\label{FEF}
\end{eqnarray}
where the maximum is taken over all possible maximally entangled states (MES). The state $\rho_{AB}$ is said to be useful for teleportation when $F(\rho_{AB}) > 2/3$, i.e., $f(\rho_{AB}) > 1/2$.

\section{Optimized protection of entanglement and TF in the presence of decoherence using the technique of WMRWM}
\label{Sec_Weak}

For the purpose of teleportation, the sender, say, Alice prepares two-qubit in one of the maximally entangled states given by
\begin{eqnarray}
|\psi\rangle_{12}^{\pm} = \frac{|00\rangle_{12} \pm |11\rangle_{12}}{\sqrt{2}}, 
\label{psi}\\
|\phi\rangle_{12}^{\pm} = \frac{|01\rangle_{12} \pm |10\rangle_{12}}{\sqrt{2}},
\label{phi}
\end{eqnarray}
Where subscript $1 (2)$ represents Alice's (Bob's) qubit. The TF and concurrence of initially prepared maximally entangled state are unity. When Alice sends the $2$nd qubit to Bob through the environment, it interacts with the environment. As a result, the shared state becomes a mixed entangled state having smaller amount of concurrence and TF~\cite{CCAA_12,YE_09,SMAHWRD_08,PCHLMK_19,PM_13,OLL_02,PCPS_07,BHHH_00, Bandyopadhyay_02}. 

To protect quantum correlation and its application in different information processing tasks, e.g., teleportation in the presence of ADC, the technique of WMRWM had been proposed~\cite{PM_13,QTYW_14,KLKK_12}. According to the technique of WMRWM, a positive-operator-valued-measurement (POVM),
\begin{eqnarray}
W_{i,0} = 
\begin{pmatrix}
1 & 0\\
0 & \sqrt{\overline{p}_i}
\end{pmatrix}
\label{WM_i}
\end{eqnarray}
has been performed on the $i$th ($i\in\{1,\,2\}$) qubit to minimize the effect of ADC. Here, $p_i$ is the strength of weak measurement, and $\overline{p}_i=1-p_i$. Weak measurement can be experimentally achieved by reducing the sensitivity of the detector, i.e., the detector never clicks if the qubit is in the state $|0\rangle_i$, and clicks with probability $p_i$ if the qubit is in the state $|1\rangle_i$~\cite{KCRK_09, LJKK_11, KLKK_12, PM_13}. Therefore, the weak measurement $W_{i,0}$ maps the initial state $\rho_i$ towards $|0\rangle_i$, which remains unaffected by ADC. After the effect of decoherence on the respective systems, a reverse weak measurement,
\begin{eqnarray}
R_{i,0} =
\begin{pmatrix}
\sqrt{\overline{q}_i} & 0 \\
0 & 1
\end{pmatrix}
\label{RWM_i}
\end{eqnarray} 
has been performed on the $i$th qubit. Here, $q_i$ is the strength of reverse weak measurement, and $\overline{q}_i=1-q_i$. Interestingly, the effect of $R_{i,0}$ is opposite to $W_{i,0}$. Finally, optimal protection of different quantum properties from ADC has been calculated from their maximization w.r.t. the parameter $q_i$. Note that weak measurements are associated with the success probability of undetected qubits by the detector. The case has been discarded when a qubit is detected either during the measurement of $W_{i,0}$ or $R_{i,0}$. Therefore, the success probability of WMRWM technique is associated with the failure of registering a qubit by the detector. In the present work, we have considered two different scenarios, e.g., {\it scenario-I} where one of the systems, say, $2$nd qubit interacts with the environment, and {\it scenario-II} where the decoherence has been applied on both qubits.\\

\noindent {\it Scenario I} : For the purpose of teleportation, Alice sends the $2$nd qubit prepared in the state of Eq.~\ref{psi}~(\ref{phi}) to Bob through the environment. To protect the quantum feature during flight, Alice makes a weak measurement $W_{2,0}$ on the $2nd$ qubit, and the combined state becomes
\begin{eqnarray}
\sigma^W_{\pm} = (I\otimes W_{2,0})\rho_{\pm}(I\otimes W_{2,0}^\dagger) \\
\overline{\sigma}^W_{\pm} = (I\otimes W_{2,0})\overline{\rho}_{\pm}(I\otimes W_{2,0}^\dagger),
\end{eqnarray}
where $\rho_{\pm}$ and $\overline{\rho}_{\pm}$ correspond to the density matrices of the states $|\psi\rangle_{12}^{\pm}$ of Eq.~(\ref{psi}) and $|\phi\rangle_{12}^{\pm}$ of Eq.~(\ref{phi}), respectively. Due to the effect of decoherence on the $2$nd qubit, Alice and Bob share either $\sigma^{\mathcal{D}}_{\pm}= U_{2,0} \sigma^W_{\pm}U_{2,0}^\dagger + U_{2,1}\sigma^W_{\pm}U_{2,1}^\dagger$ or $\overline{\sigma}^{\mathcal{D}}_{\pm}= U_{2,0} \overline{\sigma}^W_{\pm}U_{2,0}^\dagger + U_{2,1}\overline{\sigma}^W_{\pm}U_{2,1}^\dagger$, where $U_{i,j} = I\otimes \mathcal{D}_{i,j}$. After receiving the qubit, Bob makes reverse weak measurement $R_{2,0}$. Finally, they share one of the following states depending on the preparation,

\begin{eqnarray}
\sigma^R_{\pm} &=& \frac{1}{P^R_{\text Succ}}(I\otimes R_{2,0})\sigma^{\mathcal{D}}_{\pm}(I\otimes R_{2,0}^\dagger), \nonumber \\
&=& 
\begin{pmatrix}
\frac{\overline{q_2}}{\alpha} & 0 & 0 &\pm \frac{\sqrt{\overline{D_2}\,\overline{p_2}\,\overline{q_2}}}{\alpha} \\
0 & 0 & 0 & 0 \\
0 & 0 & \frac{D_2\overline{p_2}\,\overline{q_2}}{\alpha} & 0\\
\pm \frac{\sqrt{\overline{D_2}\,\overline{p_2}\,\overline{q_2}}}{\alpha} & 0 & 0 & \frac{\overline{D_2}\,\overline{p_2}}{\alpha}
\end{pmatrix},
\label{psi_R}\\
\overline{\sigma}^R_{\pm} &=& \frac{1}{P^R_{\text Succ}}(I\otimes R_{2,0})\overline{\sigma}^{\mathcal{D}}_{\pm}(I\otimes R_{2,0}^\dagger), \nonumber \\
&=& 
\begin{pmatrix}
 \frac{D_2\overline{p_2}\,\overline{q_2}}{\alpha} & 0 & 0 & 0 \\
0 & \frac{\overline{D_2}\,\overline{p_2}}{\alpha} & \pm \frac{\sqrt{\overline{D_2}\,\overline{p_2}\,\overline{q_2}}}{\alpha} & 0 \\
0 & \pm \frac{\sqrt{\overline{D_2}\,\overline{p_2}\,\overline{q_2}}}{\alpha} & \frac{\overline{q_2}}{\alpha} & 0\\
0 & 0 & 0 & 0
\end{pmatrix},
\label{phi_R}
\end{eqnarray}
where  $\alpha=(\overline{p}_2 + \overline{q}_2- D_2\overline{p_2}q_2)$. Due to the probabilistic nature of WMRWM technique, the success probability of sharing both of the states $\sigma^R_{\pm}$ and $\overline{\sigma}^R_{\pm}$ becomes
\begin{eqnarray}
P_{\text Succ}^R = Tr[(I\otimes R_{2,0})\sigma^{\mathcal{D}}_{\pm}(I\otimes R_{2,0}^\dagger)] =\frac{\alpha}{2}.
\label{P_Succ_R}
\end{eqnarray}
The FEF  and TF of both the states $\sigma^R_{\pm}$ and $\overline{\sigma}^R_{\pm}$ have same value and they are given by
\begin{eqnarray}
f_R &=& \frac{\overline{p_2}+\overline{q_2}-D_2\overline{p_2}+2 \sqrt{\overline{D_2}\,\overline{p_2}\,\overline{q}_2}}{2\alpha}, \nonumber \\
F_R&=& \frac{2f_R+1}{3},
\label{TF_W}
\end{eqnarray} 
i.e., $f_R=f(\sigma^R_{\pm})=f(\overline{\sigma}^R_{\pm})$ and $F_R=F(\sigma^R_{\pm})=F(\overline{\sigma}^R_{\pm})$.
Similarly, the concurrence of both the states becomes
\begin{eqnarray}
C_R = \frac{2\sqrt{\overline{D_2}\,\overline{p_2}\,\overline{q_2}}.}{\alpha},
\label{C_R}
\end{eqnarray}
i.e., $C_R=C(\sigma^R_{\pm})=C(\overline{\sigma}^R_{\pm})$.
Next, the optimal protection, i.e., maximization of $F_R$ and $C_R$ w.r.t. the strength of reverse weak measurement $q_2$ has been studied.\\

\noindent{\it Optimized teleportation fidelity : } To protect teleportation optimally in the presence of ADC, $F_R$ of Eq.~(\ref{TF_W}) has been maximized w.r.t. $q_2$. The optimized value of TF,
\begin{eqnarray}
F^{\max}_R = \frac{3+2\,D_2\,\overline{p_2}}{3+3\,D_2\,\overline{p_2}}
\label{TF_W_Max}
\end{eqnarray}
occurs for the choice of $q_2$ given by
\begin{eqnarray}
q_2^{\max} = \frac{3 D_2 \,\overline{p_2} + D_2^2\, \overline{p_2}^2 + p_2}{(1+D_2\overline{p_2})^2}.
\label{q2_W_TF_Max}
\end{eqnarray}
For the above choice of $q_2^{\max}$, the concurrence $C_R$ becomes
\begin{eqnarray}
C^{q_2^{\max}}_R = \frac{2}{2+D_2\,\overline{p_2}}.
\label{C_W_TF_Max}
\end{eqnarray}
Here, the success probability of Eq.~(\ref{P_Succ_R}) becomes
\begin{eqnarray}
P_{\text Succ}^{q_2^{\max}} =\frac{\overline{D_2}\,(2+D_2\,\overline{p_2})\,\overline{p_2}}{2+2\, D_2\,\overline{p_2}}.
\label{Succ_P_W_TF_Max}
\end{eqnarray}\\

\begin{figure*}[t]
\centering
\includegraphics[width=6.9 in]{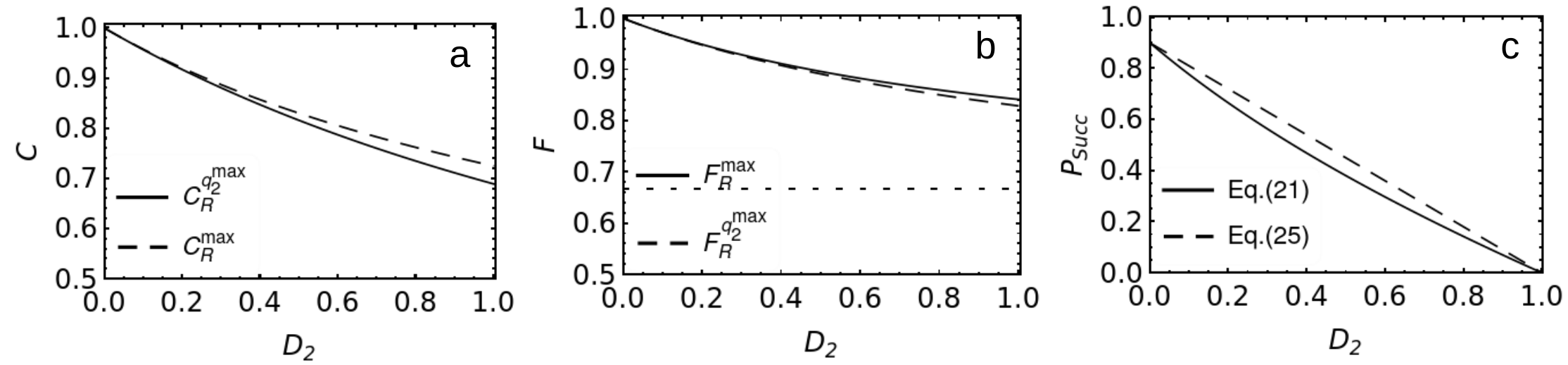}
\caption{ Comparison of improvement of (a) entanglement, (b) teleportation fidelity, (c) success probability of the states $\sigma^R_{\pm}$ (of Eq.~(\ref{psi_R})) and $\overline{\sigma}^R_{\pm}$ (of Eq.~(\ref{phi_R})) w.r.t. the strength of decoherence while considering the weak measurement strength $p_2=0.1$. The solid  and dashed lines correspond to the cases when TF and concurrence have been maximized w.r.t. $q_2$, respectively. The horizontal dashed line represents the classical upper bound of TF, $2/3$.
}

\label{Fig_Weak_2}
\end{figure*}

\noindent{\it Optimized concurrence : } Here, the concurrence $C_R$ of Eq.~(\ref{C_R}) has been maximized w.r.t. the strength of reverse weak measurement $q_2$. The maximum value of concurrence 
\begin{eqnarray}
C^{\max}_R =\frac{1}{\sqrt{1+D_2\,\overline{p_2}}}
\label{C_W_Max}
\end{eqnarray}
occurs for the 
\begin{eqnarray}
\mathfrak{q}_2^{\max} = \frac{p_2 + 2\,D_2\,\overline{p_2}}{1+D_2\,\overline{p_2}}.
\label{q2_W_C_Max}
\end{eqnarray}
For the choice of $\mathfrak{q}_2^{\max}$, the TF becomes
\begin{eqnarray}
F^{\mathfrak{q}_2^{\max}}_R = \frac{1}{6}\left( 3 + 2\frac{1}{\sqrt{1+D_2\,\overline{p_2}}} + \frac{1}{1+D_2\,\overline{p_2}}\right).
\label{TF_W_C_Max}
\end{eqnarray}
In this case, the success probability becomes
\begin{eqnarray}
P_{\text Succ}^{\mathfrak{q}_2^{\max}} =\overline{D_2}\,\overline{p_2}.
\label{Succ_P_W_C_Max}
\end{eqnarray}

The above two optimization cases have been compared in the Fig.~(\ref{Fig_Weak_2}) for the strength of weak measurement $p_2=0.1$. Figs.~\ref{Fig_Weak_2}(a)-(b) show that the strength of reverse weak measurement which optimally protects TF, does not maximize concurrence in the presence of ADC, and vice-versa. Fig.~\ref{Fig_Weak_2}(c) indicates that maximization of concurrence has a higher success probability than maximization of TF. As shown in the Fig.~(\ref{Fig_Succ_BI}) in the appendix, protection of higher form of nonlocal correlation, i.e., Bell nonlocal correlation has a lower success probability than entanglement. Therefore, the lower success probability of TF indicates that a stronger form of nonlocal correlation (stronger than entanglement) is required to get maximum TF in the presence of decoherence. Note that $\{F, \, C\} \rightarrow 1$ for $p_2 \rightarrow 1$, and the corresponding success probabilities $\{ P_{\text Succ}^{q_2^{\max}} , P_{\text Succ}^{\mathfrak{q}_2^{\max}} \} \rightarrow 0$. \\

\noindent {\it Scenario II} : Here, both qubits have been affected by the ADC, and the WMRWM technique has been applied on both qubits to protect both TF and concurrence. After preparing two qubits in the state $|\psi\rangle_{12}^{+}$ (or $|\psi\rangle_{12}^{-}$) of Eq.~(\ref{psi}), Alice makes weak measurements with strength $p_i$ on the $i$-th qubit. As a result, the combined state becomes $\sigma^{WW}_{\pm} = (W_{1,0}\otimes W_{2,0})\rho_{\pm}(W_{1,0}^\dagger \otimes W_{2,0}^\dagger)$. Due to interaction with the environment, the combined state becomes $\sigma^{\mathcal{D} \mathcal{D}}_{\pm}=V_{1,0}\sigma_{\pm}^{\mathcal{D}WW}V_{1,0}^{\dagger} + V_{1,1}\sigma_{\pm}^{\mathcal{D}WW}V_{1,1}^{\dagger}$, where $\sigma_{\pm}^{\mathcal{D}WW} = U_{2,0}\sigma_{\pm}^{WW}U_{2,0}^{\dagger} + U_{2,1}\sigma_{\pm}^{WW}U_{2,1}^{\dagger}$ and $V_{i,j}=\mathcal{D}_{i,j}\otimes I$. To minimize the environmental effect, both Alice and Bob make reverse weak measurements on their respective qubit with strength $q_i$. The shared state becomes
\begin{eqnarray}
\sigma^{RR}_{\pm}&=&\frac{1}{P_{\text Succ}^{RR}}(R_{1,0}\otimes R_{2,0})\sigma^{DD}_{\pm}(R_{1,0}^\dagger\otimes R_{2,0}^\dagger) \\
&=& 
\begin{pmatrix}
\frac{\overline{q}^2(1+D^2\overline{p}^2)}{\beta} & 0 & 0 & \frac{(-1)^{1+m}\overline{D}\,\overline{p}\,\overline{q}}{\beta} \\
0 & \frac{D\,\overline{D}\,\overline{p}^2\,\overline{q}}{\beta} & 0 & 0 \\
0 & 0 & \frac{D\,\overline{D}\,\overline{p}^2\,\overline{q}}{\beta} & 0\\
\frac{(-1)^{1+m}\overline{D}\,\overline{p}\,\overline{q}}{\beta} & 0 & 0 & \frac{\overline{D}^2\,\overline{p}^2}{\beta}
\end{pmatrix}, \nonumber
\label{State_RR}
\end{eqnarray}
where $\beta=2-2 q (1+D\,\overline{p}^2) + q^2(1+D^2\,\overline{p}^2)-(2-p)p$. Here, for simplicity, $D_2=D_1=D$, $p_2=p_1=p$ and $q_2=q_1=q$ have been considered. Similar to the {\it Scenario I}, here, two different cases, i.e., {\it optimized teleportation fidelity} and {\it optimized concurrence}, have been considered. The success probability of getting the state $\sigma^{RR}_{\pm}$ becomes
\begin{eqnarray}
P_{\text Succ}^{RR} =&&\frac{\beta}{2}
\end{eqnarray}
For the shared state $\sigma^{RR}_{\pm}$, the FEF becomes
\begin{eqnarray}
f_{RR} = \frac{D^2(1+\overline{q}^2)\overline{p}^2-2D\overline{p}(\overline{p}+\overline{q}) + (\overline{p}+\overline{q})^2}{\beta},
\label{f_WW}
\end{eqnarray}
and the concurrence is given below
\begin{eqnarray}
C_{RR} = \frac{\overline{D}\,\overline{p}\,\overline{q}( \delta_1 - \delta_2 -2D\overline{p})}{\beta},
\end{eqnarray}
where $\delta_1=\sqrt{2(1+\sqrt{1+D^2\overline{p}^2}) + D^2\overline{p}^2}$ and $\delta_2=\sqrt{2(1-\sqrt{1+D^2\overline{p}^2}) + D^2\overline{p}^2}$. \\

\noindent {\it Optimized teleportation fidelity : } In this scenario, the optimized TF becomes 
\begin{eqnarray}
F^{\max}_{RR} = \frac{ \left(2+( 1 - \eta)\left(\sqrt{1+\eta^2} - \eta \right)\right) }{3},
\label{TF_WW_TF_Max}
\end{eqnarray}
where $\eta=D\overline{p}$, and it is obtained  for the choice of the strength of reverse weak measurement
\begin{eqnarray}
q^{\max} =1-\frac{\overline{p}\overline{D}}{\sqrt{1+\eta^2}}.
\label{q_WW_TF_Max} 
\end{eqnarray}
For the above choice of $q^{\max}$, the concurrence becomes 
\begin{eqnarray}
C^{q^{\max}}_{RR} = \frac{1}{2}\left( \sqrt{1+\eta^2}-\eta \right)\left(\delta_1-\delta_2 - 2D\overline{p} \right).
\label{C_WW_TF_Max}
\end{eqnarray}

\noindent{\it Optimized concurrence : } Interestingly, in this case, optimized protection of TF implies optimized protection of concurrence in the presence of ADC. Therefore the maximum value of concurrence occurs for the same strength of reverse weak measurement $q^{\max}$ of Eq.~(\ref{q_WW_TF_Max}). The maximum value of concurrence becomes $C^{\max}_{RR}=C^{q^{\max}}_{RR}$, and the corresponding TF has the form of $F^{\max}_{RR}$.

Note here that when both qubits are affected by ADC, the optimum protection of quantum feature using the WMRWM technique fails for the prepared state in the form of $|\phi\rangle_{12}^{\pm}$ of Eq.~(\ref{phi}). The weak measurement by Alice on both qubits does not affect the state.

\section{Conclusion}
In the present work, we have studied optimized protection of quantum features, e.g., entanglement and teleportation fidelity in the presence of ADC, using the technique of WMRWM. We have found that when both qubits interact with the environment, the optimized protection of TF and entanglement occur for the same value of the strength of reverse weak measurement. More specifically, here, optimization of TF implies optimized protection of concurrence. Interestingly this phenomenon does not hold when one of the qubits interacts with the environment via ADC, i.e., optimized protection of entanglement does not maximize the TF. Although entanglement is necessary to have non-classical TF, but, the success probability (as shown in the Fig.~(\ref{Fig_Succ_BI})) reveals that optimum TF requires stronger form of nonlocal correlation than entanglement. This opens the question of general resource of teleportation for further studies.

\section{Data availability statement}
No new data were created or analysed in this study


\appendix
\section{Protection of  Bell nonlocal correlation}
\label{Apdx_Steer}
Here, we have studied the optimal protection of Bell nonlocal correlation in the presence of decoherence for the {\it Scenario I}. Bell violation of the state $\sigma^R_{\pm}$ of Eq.~(\ref{psi_R}) ($\overline{\sigma}^R_{\pm}$ of Eq.~(\ref{phi_R})) can be calculated from the correlation matrix $\mathcal{T} = \{\mathcal{T}_{i,j}=Tr[(\sigma_i\otimes \sigma_j)\sigma^R_{\pm}]\}$~\cite{HHH_95, ZC_02}. The Bell violation of the state $\sigma^R_{\pm}$ ($\overline{\sigma}^R_{\pm}$)  is given by
\begin{eqnarray}
BI &=& 2\sqrt{\lambda_1 + \lambda_2}, \nonumber \\
&=& \frac{4\overline{D}_2 \overline{p}_2 \overline{q}_2 + (D_2(1+\overline{q}_2) \overline{p}_2 - \overline{p}_2 - \overline{q}_2)^2}{\overline{p}_2+\overline{q}_2-D_2 \overline{p}_2 q_2  }
\end{eqnarray}
where $\lambda_1$ and $\lambda_2$ are two maximum eigenvalues of $\mathcal{T}^{T}\mathcal{T}$ (here, superscript $T$ stands for the transposition). The maximized value of $BI$ occurs for $q_2=\frac{D_2 (p_2-1) (2 D_2 (p_2-1)-p_2+4)-p_2}{D_2(p_2-1) (D2 (p_2-1)+2)-1}$, and the corresponding success probability becomes 

\begin{figure}[t]
\centering
\includegraphics[width=3.5 in]{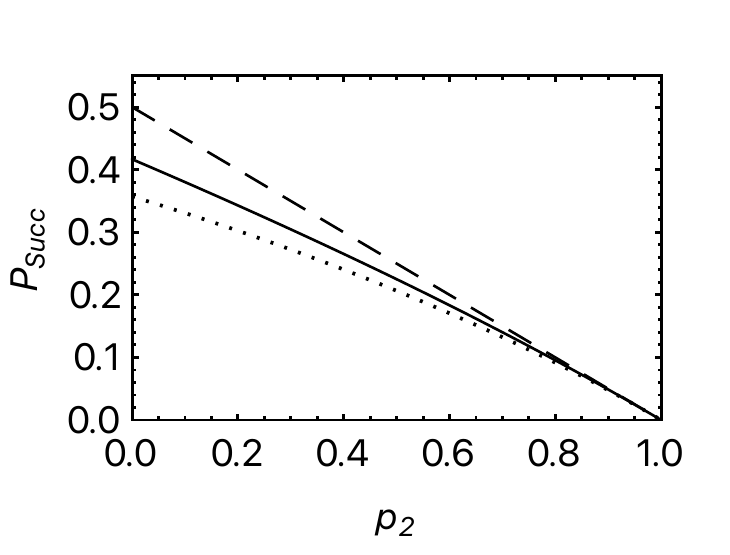}
\caption{ Comparison of success probability of the states $\sigma^R_{\pm}$ (of Eq.~(\ref{psi_R})) and $\overline{\sigma}^R_{\pm}$ (of Eq.~(\ref{phi_R})) w.r.t. the strength of weak measurement, $p_2$ while considering the decoherence strength $D_2=0.5$. The dashed, solid, and dotted lines correspond to $P_{\text Succ}^{\mathfrak{q}_2^{\max}}$ (of Eq.~(\ref{Succ_P_W_C_Max})), $P_{\text Succ}^{q_2^{\max}}$ (of Eq.~ (\ref{Succ_P_W_TF_Max})), $P_{\text{Succ}}^{BI}$ (of Eq.~(\ref{Succ_BI_Max})), respectively. 
}
\label{Fig_Succ_BI}
\end{figure}
\begin{eqnarray}
P_{\text{Succ}}^{BI}= \frac{\overline{D}_2 \overline{p}_2 (1+D_2\overline{p}_2 - D_2^2\overline{p}_2^2)}{1+2D_2\overline{p}_2+D_2^2\overline{p}_2^2}.
\label{Succ_BI_Max}
\end{eqnarray}
The Fig.~(\ref{Fig_Succ_BI}) shows the comparison of success probabilities when concurrence, TF and Bell violation are maximized w.r.t. the strength of reverse weak measurement $q_2$. It reflects that a stronger form of nonlocal correlation has a lower success probability.


\begin{thebibliography}{99}


\bibitem{BBCJP_93}   Bennett C H, Brassard G, Cr\'{e}peau C, Jozsa R, Peres A, and Wootters W K 1993 Phys. Rev. Lett. {\bf 70} 1895.

\bibitem{BPMEWZ_97} Bouwmeester D, Pan J W, Mattle  K, Eibl M, Weinfurter H, and Zeilinger A 1997 Nature {\bf 390} 575.

\bibitem{DBMHP_98} Boschi D, Branca S, Martini F D, Hardy L, and Popescu S 1998 Phys. Rev. Lett. {\bf 80} 1121.

\bibitem{MRTZG_03}  Marcikic I, Riedmatten  H D, Tittel W, Zbinden H, and Gisin N 2003 Nature {\bf 421}, 509.

\bibitem{Ren_al_17} Ren J. G. et al 2017 Nature {\bf 549} 70.

\bibitem{Book_NC} Nielsen M A and Chuang I L 2000 {\it Quantum computation and quantum information} (Cambridge University Press).


\bibitem{Ent_rev} Horodecki R, Horodecki P, Horodecki M, and Horodecki K 2009 Rev. Mod. Phys. {\bf 81} 865.

\bibitem{PW_91} Peres A, Wootters W K 1991 Phys. Rev. Lett. {\bf 66} 1119.

\bibitem{MP_95} Massar S, Popescu S 1995 Phys. Rev. Lett. {\bf 74} 1259.


\bibitem{JY_12}  Yin J et al 2012 Nature {\bf 488} 185.

\bibitem{XSM_12}  Ma X S et al 2012  Nature {\bf 489} 269.


\bibitem{CCAA_12} Chaves R, Cavalcanti D, Aolita L, and Ac\'{i}n A 2012 Phys. Rev. A {\bf 86} 012108.

\bibitem{YE_09}  Yu T and Eberly J H 2009 Science {\bf 323} 598.

\bibitem{SMAHWRD_08} Salles A, Melo F D, Almeida M P, Hor-Meyll M, Walborn S P,  Souto Ribeiro P H, and Davidovich  L 2008 Phys. Rev. A {\bf 78} 022322.

\bibitem{AMHSWRD_07} Almeida M P, Melo F D, Hor-Meyll M, Salles A, Walborn S P, Ribeiro P H S, Davidovich L 2007 Science {\bf 316} 579.

\bibitem{PCHLMK_19} Pramanik T, Cho Y W, Han  S W, Lee S Y, Moon  S, and Kim Y S 2019 Phys. Rev. A {\bf 100} 042311.

\bibitem{PM_13}  Pramanik T, Majumdar A S 2013 Phys. Lett. A {\bf 377} 3209.

\bibitem{OLL_02} Oh S, Lee S, and Lee H W 2002 Phys. Rev. A {\bf 66} 022316.

\bibitem{PCPS_07} Prakash H, Chandra N, Prakash R, and Shivani 2007 J. Phys. B: At. Mol. Opt. Phys. {\bf 40} 1613.

\bibitem{PH_02} Plenio M B, Huelga  S F 2002 Phys. Rev. Lett. {\bf 88} 197901.

\bibitem{DB_02}  Braun D 2002 Phys. Rev. Lett. {\bf 89} 277901.

\bibitem{KLAK_02} Kim  M S, Lee J, Ahn D, and Knight P L 2002 Phys. Rev. A {\bf 65} 040101(R).

\bibitem{HZC_23} Harraz  S, Zhang J-Y, Cong S 2023 arXiv:2206.14463.

\bibitem{SKB_23} Sabale V B, Kumar A, Banerjee S 2023 arXiv:2307.09231 

\bibitem{KLKK_12} Kim Y S,  Lee J C, Kwon O, Kim Y H 2012 Nat. Phys. {\bf 8} 117.

\bibitem{LHK_15} Lim H T, Hong K H, and Kim Y H 2015 Sci. Rep. {\bf 5} 15384.

\bibitem{KU_99} Koashi M, Ueda M 1999 Phys. Rev. Lett. {\bf 82} 2598.

\bibitem{KCRK_09} Kim Y S, Cho  Y W,  Ra Y S, Kim Y H 2009 Opt. Express {\bf 17} 11978.

\bibitem{BHHH_00} Badziag  P, Horodecki M, Horodecki P, Horodecki R 2000 Phys. Rev. A {\bf 62} 012311.

\bibitem{Bandyopadhyay_02} Bandyopadhyay S 2002 Phys. Rev. A {\bf 65} 022302.

\bibitem{NDDA_18} Nandi S, Datta C, Das A, and Agrawal  P 2018   Eur. Phys. J. D {\bf 72} 182.

\bibitem{QTYW_14} Qiu L, Tang G, Yang  X, and Wang A 2014 Annals of Physics, {\bf 350}, 137.

\bibitem{SP_95} Popescu S 1995 Phys. Rev. Lett. {\bf 74} 2619.

\bibitem{PHBK_99} Plenio M B, Huelga S F, Beige A, and Knight P L 1999 Phys. Rev. A {\bf 59} 2468.

\bibitem{NV_11} Navasc\'{u}es M, and Ver\'{t}esi T 2011 Phys. Rev. Lett. {\bf 106} 060403.

\bibitem{PCHLKM_19} Pramanik T, Cho Y W, Han S W, Lee S Y, Kim Y S, and Moon S 2019 Phys. Rev. A {\bf 99}, 030101(R).

\bibitem{LJKK_11} Lee J C, Jeong Y C, Kim Y S, Kim Y H 2011 Opt. Express {\bf 19} 16309.


\bibitem{Wootters_98} Wootters W K 1998 Phys. Rev. Lett. {\bf 80}, 2245.

\bibitem{HHH_99} Horodecki M, Horodecki P, Horodecki R 1999 Phys. Rev. A {\bf 60} 1888.

\bibitem{BVSW_96} Bennett C H, Vincenzo D P D, Smolin J A, and Wootters  W K 1996 Phys. Rev. A {\bf 54} 3824.







\bibitem{HHH_95} Horodecki R, Horodecki P and Horodecki M 1995 Phys. Lett. A {\bf 200} 340.

\bibitem{ZC_02} Zukowski M and Brukner C 2002 Phys. Rev. Lett. Phys. Rev. Lett. {\bf 88} 210401.








\end{thebibliography}
\end{document}